\documentclass[amsmath,floatfix,twocolumn,superscriptaddress]{revtex4-1}
\usepackage{subfigure}
\usepackage{siunitx}
\usepackage{amssymb}
\usepackage{amsmath}
\usepackage{graphicx}
\usepackage{array}
\usepackage{dcolumn}
\usepackage{psfrag}
\usepackage{bm}
\usepackage{color}
\usepackage{multirow}

\begin{document}

\title{Band-Gap Tunability in Anharmonic Perovskite-like Semiconductors Driven by Polar Electron-Phonon Coupling}

\author{Pol Benítez}
\email{pol.benitez@upc.edu}
    \affiliation{Department of Physics, Universitat Politècnica de Catalunya, 08034 Barcelona, Spain}
    \affiliation{Research Center in Multiscale Science and Engineering, Universitat Politècnica de Catalunya, 
    08019 Barcelona, Spain}

\author{Ruoshi Jiang}
    \affiliation{Department of Materials Science and Metallurgy, University of Cambridge, Cambridge CB3 0FS, UK}

\author{Siyu Chen}
    \affiliation{Department of Materials Science and Metallurgy, University of Cambridge, Cambridge CB3 0FS, UK}

\author{Cibrán López}
    \affiliation{Department of Physics, Universitat Politècnica de Catalunya, 08034 Barcelona, Spain}
    \affiliation{Research Center in Multiscale Science and Engineering, Universitat Politècnica de Catalunya, 
    08019 Barcelona, Spain}

\author{Josep-Lluís Tamarit}
    \affiliation{Department of Physics, Universitat Politècnica de Catalunya, 08034 Barcelona, Spain}
    \affiliation{Research Center in Multiscale Science and Engineering, Universitat Politècnica de Catalunya, 
    08019 Barcelona, Spain}

\author{Edgardo Saucedo}
    \affiliation{Research Center in Multiscale Science and Engineering, Universitat Politècnica de Catalunya, 
    08019 Barcelona, Spain}
    \affiliation{Department of Electronic Engineering, Universitat Politècnica de Catalunya, 08034 Barcelona, Spain}

\author{Bartomeu Monserrat}
\email{bm418@cam.ac.uk}
    \affiliation{Department of Materials Science and Metallurgy, University of Cambridge, Cambridge CB3 0FS, UK}

\author{Claudio Cazorla}
\email{claudio.cazorla@upc.edu}
    \affiliation{Department of Physics, Universitat Politècnica de Catalunya, 08034 Barcelona, Spain}
    \affiliation{Research Center in Multiscale Science and Engineering, Universitat Politècnica de Catalunya, 
    08019 Barcelona, Spain}

\begin{abstract}
\textbf{Abstract.}~The ability to finely tune optoelectronic properties in semiconductors is crucial for the development of advanced technologies, ranging from photodetectors to photovoltaics. In this work, we propose a novel strategy to achieve such tunability by utilizing electric fields to excite low-energy polar optical phonon modes, which strongly couple to electronic states in anharmonic semiconductors. We conducted a high-throughput screening of over $10,000$ materials, focusing on centrosymmetric compounds with imaginary polar phonon modes and suitable band gaps, and identified $310$ promising candidates with potential for enhanced optoelectronic tunability. From this set, three perovskite-like compounds --Ag$_3$SBr, BaTiO$_3$, and PbHfO$_3$-- were selected for in-depth investigation based on their contrasting band-gap behavior with temperature. Using first-principles calculations, \textit{ab initio} molecular dynamics simulations, tight-binding models, and anharmonic Fröhlich theory, we analyzed the underlying physical mechanisms. Our results show that polar phonon distortions can induce substantial band-gap modulations at ambient conditions, including reductions of up to $70\%$ in Ag$_3$SBr and increases of nearly $23\%$ in BaTiO$_3$, relative to values calculated at zero temperature, while PbHfO$_3$ exhibits minimal change. These contrasting responses arise from distinct electron-phonon coupling mechanisms and orbital hybridization at the band edges. This work establishes key design principles for harnessing polar lattice dynamics to engineer tunable optoelectronic properties, paving the way for adaptive technologies such as wavelength-selective optical devices and solar absorbers.
\\

{\bf Keywords:} electron-phonon coupling, first-principles calculations, anharmonicity, perovskite materials, optoelectronic properties
\end{abstract}

\maketitle

\section*{Introduction}
\label{sec:intro}
Semiconductors are foundational to modern technologies and play a critical role in a wide range of applications, including optoelectronics, computing, information storage, and energy harvesting. A defining feature of these materials is the electronic band gap, namely, the energy difference between the valence band maximum (VBM) and the conduction band minimum (CBM), which typically spans a few eV. Achieving significant externally controlled band-gap variations in semiconductors is crucial for enabling tunable electronic and optoelectronic devices, such as transistors, sensors, and photodetectors.

The band gap of semiconductors is known to vary with temperature \cite{monserrat2017temperature,monserrat2018role}, primarily due to electron-phonon interactions \cite{giustino2017electron} and thermal expansion of the crystal lattice \cite{malloy1991thermal}. While these variations are typically modest, on the order of a few meV \cite{bube1955temperature,bludau1974temperature,wu2003temperature}, certain materials exhibit much larger shifts, reaching several hundred meV at room temperature \cite{monserrat2015giant,benitez2025giant}. For most compounds, the band gap decreases with increasing temperature, a trend well described by the empirical Varshni equation \cite{varshni1967temperature}. However, \textit{anomalous} cases also exist \cite{villegas2016anomalous}, where the band gap increases with temperature. This behavior can manifest in two forms: a non-monotonic dependence, in which the band gap initially increases and then decreases, and a monotonic increase over the entire temperature range. Representative examples include the chalcopyrite ZnSnAs$_2$ \cite{bhosale2012temperature} (non-monotonic) and the copper halide CuCl \cite{gobel1998effects} (monotonic).

Although temperature can influence the optoelectronic properties of semiconductors through mechanisms such as electron-phonon interactions and thermal expansion, it is inherently limited as a tool for dynamic control. In practical applications, such as tunable photodetectors, adaptive photovoltaics, or reconfigurable optoelectronic devices, one requires fast, reversible, and spatially controlled modulation of material properties. Temperature changes are typically slow, energetically inefficient, and lack spatial precision. A more practical and versatile approach would involve applying external stimuli, such as electric fields, that can dynamically and selectively manipulate the electronic structure. In this context, identifying mechanisms by which external fields can induce band gap modifications, particularly through their interaction with the lattice, is of critical importance for the development of next-generation functional materials.

On a more fundamental level, despite the availability of extensive empirical data on temperature-dependent band gap shifts, predictive frameworks for anticipating such behavior in unexplored materials remain limited. Simple and chemically intuitive descriptors, such as elemental composition, bonding characteristics, or crystal symmetry, often fail to reliably predict whether a material will exhibit significant, negligible, or anomalous band gap variations with temperature. A potentially crucial factor in this context is lattice anharmonicity \cite{zhao2025distinguishing,monserrat2013anharmonic}. 

Anharmonic lattice dynamics, characterized by low-energy and large-amplitude phonon modes at the harmonic level, can strongly modulate electron-phonon interactions, thereby amplifying temperature-induced band gap shifts. However, the microscopic mechanisms linking anharmonicity to electronic structure renormalization remain poorly understood and are seldom captured by conventional theoretical models. Few recent studies have begun to explore this connection in specific materials, such as SrTiO$_3$ \cite{Wu2020} and CuInTe$_2$ \cite{yu2022}, providing valuable insights into how anharmonic effects might be harnessed to design materials with tunable optoelectronic properties.

In this work, we propose a set of electronic and lattice vibrational criteria to identify materials with the potential for large band-gap variations, whether driven by temperature or external electric fields. Based on these criteria, we perform a high-throughput computational screening of a large materials database comprising thousands of precomputed electronic band structures and phonon spectra, ultimately identifying over $300$ promising candidates. From this set, we select three highly anharmonic perovskite-like compounds, namely, Ag$_{3}$SBr, BaTiO$_{3}$ and PbHfO$_{3}$, for in-depth investigation using a combination of first-principles methods, including density functional theory (DFT), \textit{ab initio} molecular dynamics (AIMD), tight-binding (TB) models, and anharmonic Fröhlich theory. Our theoretical predictions for the room-temperature renormalized band gaps of these three compounds show excellent agreement with experimental data, thereby validating the robustness of the proposed approach.

Our analysis uncovers a set of simple, chemically intuitive mechanisms that account for the observed trends in temperature-induced band-gap variations. In particular, strong electron-phonon coupling mediated by low-frequency polar phonon modes can give rise to either conventional Varshni-like behavior or anomalous temperature dependencies, depending on the specific orbital hybridizations at the band edges. These findings demonstrate the feasibility of rationally designing optoelectronic materials whose properties can be tuned through phonon-mediated interactions, potentially enhanced by external electric fields, thereby 
opening avenues for innovative technologies.

\begin{figure*}
    \centering
    \includegraphics[width=1.0\linewidth]{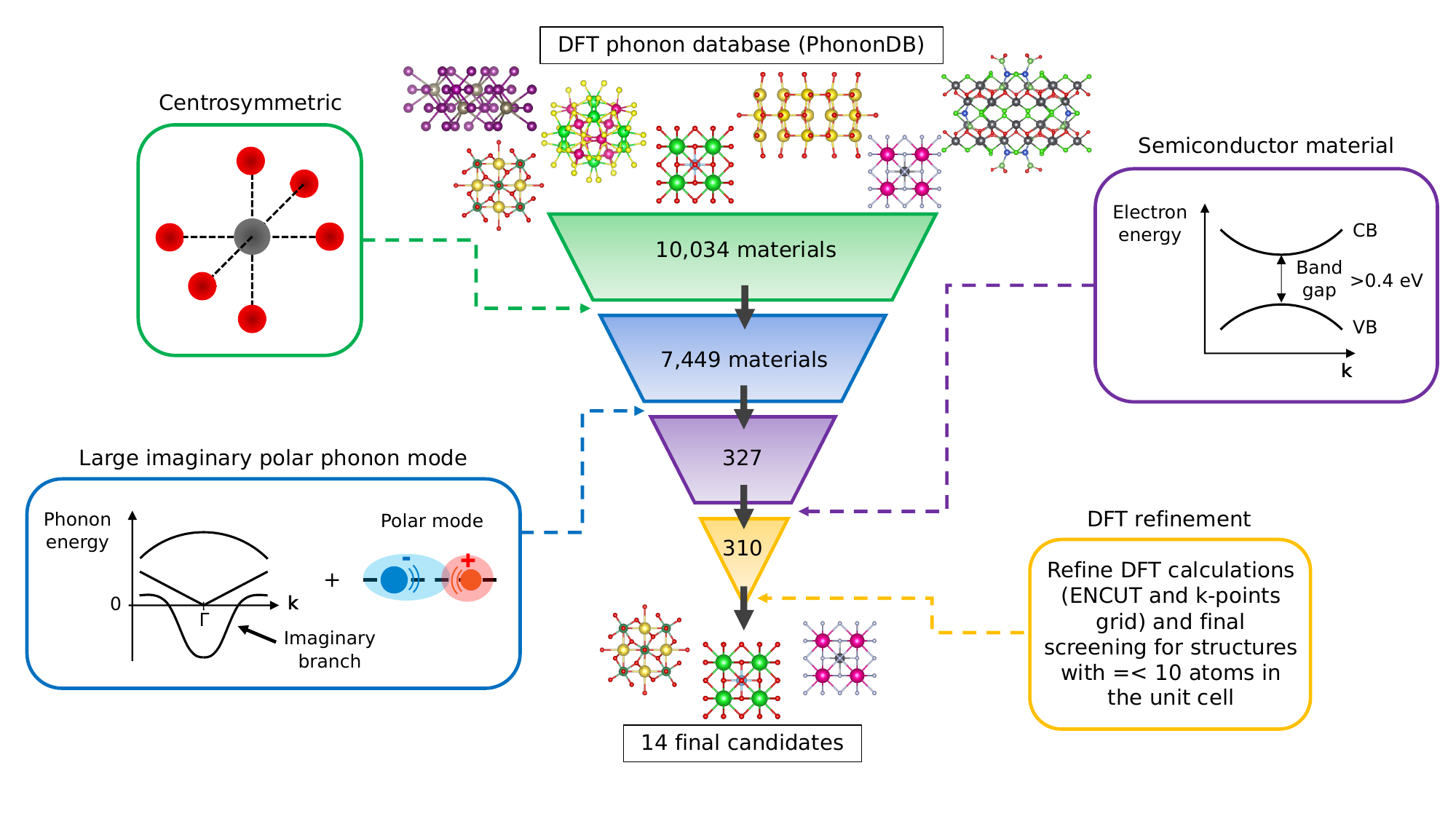}
    \caption{\textbf{Representation of the adopted materials screening strategy.}
	The screening process started with $10,034$ materials and ended up with $14$ final candidates. Structures were filtered by their crystal symmetry, $\Gamma$ phonon modes, and band gap. Refined DFT calculations were conducted for $310$ candidates with $10$ atoms or less per unit cell. The atomic structure of some illustrative materials are represented.} 
    \label{fig1}
\end{figure*}

\section*{Results}
\label{sec:results}
We begin this section by detailing the criteria used for the high-throughput screening of semiconductor materials with potentially high optoelectronic tunability, along with the rationale behind these choices. We then discuss the most promising candidates identified in our search and present a refinement of the precomputed first-principles data associated with them. Based on band-gap calculations performed on frozen-phonon distorted configurations, we select three representative compounds, all exhibiting perovskite-like structures although contrasting temperature-dependent band-gap behaviors, for a detailed investigation. Finally, we elucidate the underlying electron-phonon coupling and orbital hybridization mechanisms responsible for the observed trends, offering a unified and chemically intuitive framework to interpret the results.

\subsection*{High-throughput screening of tunable optoelectronic materials}
\label{subsec:high-throughput}
A high-throughput screening of crystalline materials was performed using the computational phonon database PhononDB \cite{phonondb}. This database contains phonon calculations based on finite-displacement methods carried out with the \verb!PhonoPy! software package \cite{phonopy-phono3py-JPCM,phonopy-phono3py-JPSJ}. The database includes $10,034$ distinct crystal structures, all of which originate from The Materials Project \cite{jain2013commentary}, thereby enabling straightforward access to complementary material properties such as electronic band gaps.

\begin{figure*}
    \centering
    \includegraphics[width=0.9\linewidth]{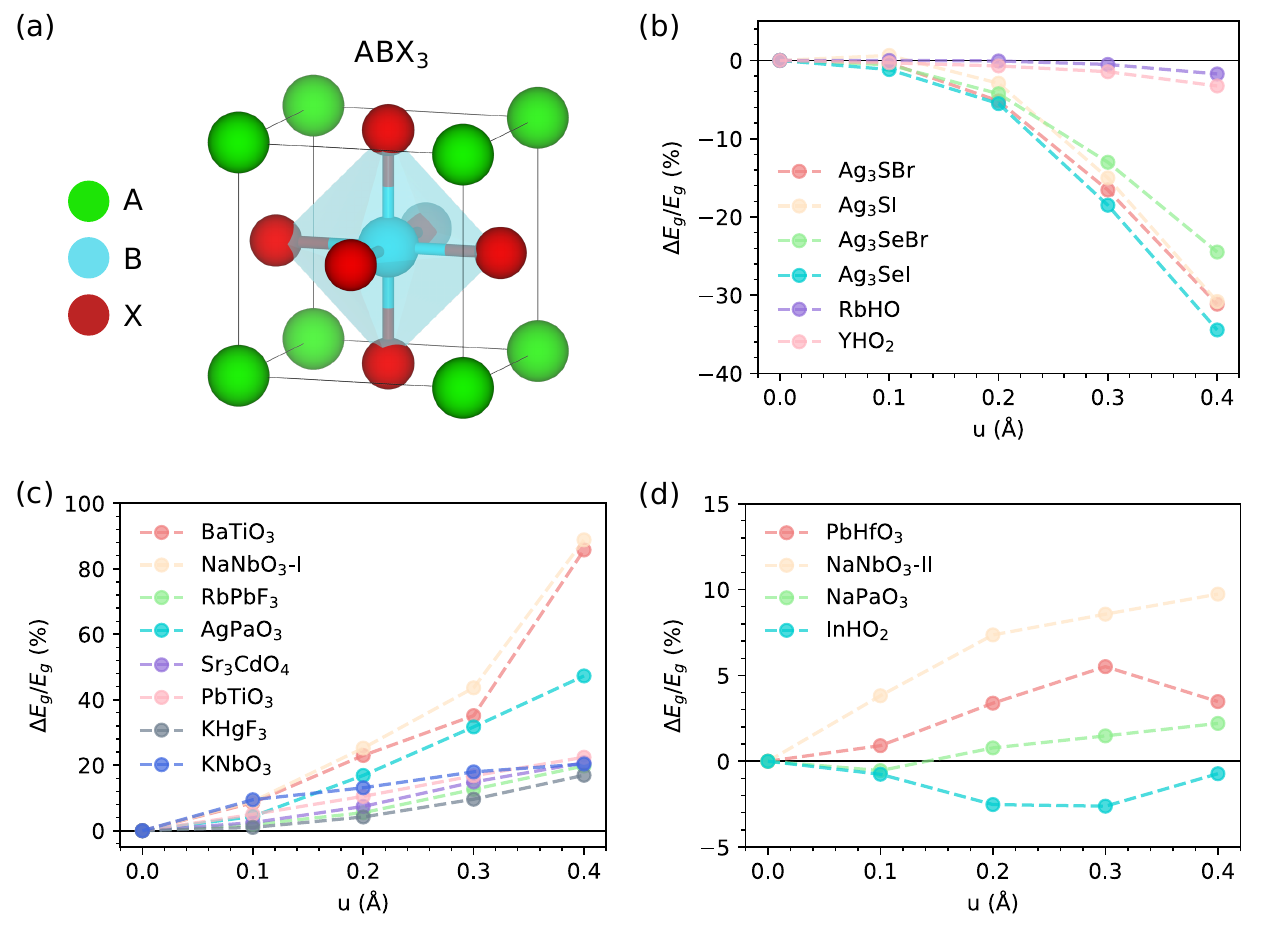}
    \caption{\textbf{Perovskite-like structure and relative variation of the band gap under frozen-phonon distortions.}(a)~Perovskite and anti-perovskite structures: B atoms are cations in perovskites and anions in anti-perovskites; X atoms are anions in perovskites and cations in anti-perovskites. (b--d)~Candidate materials grouped by phonon-driven band-gap variation: reduction, increase, and minimal change. NaNbO$_3$-I and II correspond to two different polymorphs with $Pm\overline{3}m$ and $P4/mbm$ symmetry, respectively.}
    \label{fig2}
\end{figure*}

Figure~\ref{fig1} outlines the screening workflow and materials selection criteria adopted in this study. The first step involved identifying those with centrosymmetric crystal structures, that is, structures possessing inversion symmetry. The rationale for focusing on centrosymmetric materials was twofold. First, introducing phonon-like distortions in these highly symmetric structures can lift electronic band degeneracies, a well-known mechanism for modifying band gaps \cite{berger2011band}. Second, centrosymmetric systems can host polar phonon distortions that break inversion symmetry and may be externally activated by electric fields. This may allow for controlled and efficient tuning of optoelectronic properties via field-induced lattice distortions. A total of $7,449$ materials met this symmetry criterion and advanced to the next stage of the screening process.

As stated in the Introduction, our focus is on anharmonic materials due to their tendency to host low-energy optical phonon modes. Such modes typically involve large atomic displacements and are likely to induce strong electron-phonon coupling. Although the PhononDB database contains only harmonic phonon calculations, it is still possible to infer signatures of anharmonicity from them. One such indicator is the presence of imaginary phonon modes. Indeed, strongly anharmonic materials, while dynamically stable at finite temperatures, may exhibit imaginary frequencies in phonon dispersion relations calculated at zero temperature \cite{tolborg2023exploring}. Based on this rationale, our next screening criterion selected materials with at least one imaginary phonon frequency below $1.5$i~THz at the $\Gamma$-point. This frequency threshold was chosen to avoid the inclusion of spurious imaginary modes arising from the neglect of long-range multipolar interactions \cite{royo2020using}. For simplicity, we restricted our analysis to $\Gamma$-point phonons, although materials exhibiting imaginary modes elsewhere in the Brillouin zone may be also promising.

Additionally, we required the aforementioned imaginary $\Gamma$-point phonon mode to be polar, that is, it should involve a net displacement between positively and negatively charged ions, thereby generating an electric dipole \cite{gonze1997dynamical}. Such polar modes are capable of coupling to external electric fields, hence are essential for applications \cite{baroni2001phonons}. Applying this additional criterion narrowed the list to $327$ candidate materials. We then further reduced the dataset by retaining only semiconductor materials with a non-zero band gap. Band gap values were obtained from The Materials Project \cite{jain2013commentary}, and materials with band gaps smaller than $0.4$~eV were excluded to ensure well-defined semiconductor behavior. Of the initial $327$ candidates, $310$ compounds met this criterion. These materials, along with their Materials Project ID, chemical formula, space group, number of atoms in the unit cell, and band gap, are listed in the Supplementary Files. 

To refine those results with more accurate first-principles calculations, we recomputed selected properties using higher-precision parameters than those employed in PhononDB \cite{phonondb} and The Materials Project \cite{jain2013commentary} (Methods). However, many of the shortlisted structures contained a large number of atoms per unit cell, ranging from dozens to hundreds, making such recalculations computationally very demanding. To ensure feasibility, we restricted this final refinement step to structures with ten or fewer atoms per unit cell, resulting in a subset of $24$ candidates. For these, we performed higher-accuracy geometry relaxations and $\Gamma$-point phonon calculations using DFT. The updated values are provided in Supplementary Table~I. Of the $24$ materials, $14$ continued to meet all the previous screening criteria following this refinement.

Among the $14$ shortlisted candidates, $10$ exhibit perovskite-like structures with the general formula ABX$_3$ and adopt the typical paraelectric high-temperature cubic phase with space group $Pm\overline{3}m$ \cite{vogt1993high}, illustrated in Fig.~\ref{fig2}a. The remaining four materials include one compound with a perovskite-related structure in the $P4/mbm$ space group and three ternary compounds with $P2_{1}/m$ and $Pnnm$ symmetries. To this list, we added four chalcohalide antiperovskite compounds, namely, Ag$_3$SBr, Ag$_3$SI, Ag$_3$SeBr, and Ag$_3$SeI, previously investigated in works \cite{benitez2025giant,cano2024novel,benitez2025crystal} although not present in the PhononDB database. These compounds were considered \textit{a posteriori} due to their marked anharmonic behavior and strong electron-phonon coupling \cite{benitez2025giant}. Like the ABX$_3$ perovskites, they adopt a cubic structure with $Pm\overline{3}m$ symmetry at finite temperatures \cite{cano2024novel,benitez2025crystal}. This addition brings the total number of analyzed materials to $18$. Table~I summarizes these candidates, reporting their chemical composition, space group, energy of the most unstable (imaginary) phonon mode, and band gap calculated at zero temperature.

\begin{table}[!htbp]
    \centering
    \begin{tabular}{cccc}
    \hline
    \hline
	    & & &    \\
	    \quad Material \quad & \quad Space group \quad & $ \quad E_{\Gamma}$ \quad & $ \quad E_{g}^{0{\rm K}}$ \qquad  \\
	                   &                         &          \quad  (meV)               &   \quad       (eV)           \\
	    & & &  \\
    \hline
	    & & & \\
            Ag$_3$SBr 		& $Pm\overline{3}m$ 	& 9.00i 	& 1.8 \\
            Ag$_3$SI 		& $Pm\overline{3}m$ 	& 9.12i 	& 1.4 \\
            Ag$_3$SeBr 		& $Pm\overline{3}m$ 	& 9.57i 	& 1.6 \\
            Ag$_3$SeI 		& $Pm\overline{3}m$ 	& 9.16i 	& 1.3 \\
            PbHfO$_3$ 		& $Pm\overline{3}m$ 	& 15.09i 	& 3.2 \\
            PbTiO$_3$ 		& $Pm\overline{3}m$ 	& 14.66i 	& 2.3 \\
	    BaTiO$_3$ 		& $Pm\overline{3}m$ 	& 16.75i 	& 2.5 \\
            KNbO$_3$ 		& $Pm\overline{3}m$ 	& 24.43i 	& 2.4 \\
            NaNbO$_3$-I 	& $Pm\overline{3}m$ 	& 21.36i 	& 2.5 \\
            NaNbO$_3$-II 	& $P4/mbm$ 		& 18.88i 	& 2.5 \\
            KHgF$_3$ 		& $Pm\overline{3}m$ 	& 8.39i 	& 1.8 \\
            RbPbF$_3$ 		& $Pm\overline{3}m$ 	& 10.41i 	& 3.6 \\
            NaPaO$_3$ 		& $Pm\overline{3}m$ 	& 11.24i 	& 4.5 \\
            AgPaO$_3$ 		& $Pm\overline{3}m$ 	& 8.85i 	& 1.5 \\
            Sr$_3$CdO$_4$ 	& $Pm\overline{3}m$ 	& 6.05i 	& 1.6 \\
            YHO$_2$ 		& $P2_1/m$ 		& 23.17i 	& 5.5 \\
            InHO$_2$ 		& $Pnnm$ 		& 20.36i 	& 3.2 \\
            RbHO 		& $P2_1/m$ 		& 41.95i 	& 4.7 \\
            & & &   \\
    \hline
    \hline
    \end{tabular}
	\caption{{\bf Selected materials based on our high-throughput screening.} 
	Summary of the structural, vibrational and band-gap properties of the $18$ final candidates. 
	$E_{\Gamma}$ represents the (imaginary) energy of the largest $\Gamma-$point polar phonon 
	instability, and $E_{g}^{0{\rm K}}$ the band gap calculated at zero temperature.}
    \label{tab:materials}
\end{table}

\subsection*{Band gap change induced by low-energy optical polar phonon displacements}
\label{subsec:bangap-change}
For each of the $18$ sieved materials, we applied unit-cell structural distortions along the eigenvectors of their most unstable polar $\Gamma$ phonon modes (i.e., those with the largest imaginary frequencies). The total physical reasonable displacement amplitudes ranged from zero to $0.4$~\AA, and for each distorted configuration we computed the band gap, $E_{g}$ (Methods). The relative percentage change in $E_{g}$ as a function of the distortion amplitude is presented in Figs.~\ref{fig2}b--d. Based on these results, the materials were classified into three categories: those exhibiting a monotonic decrease in $E_{g}$ (Fig.~\ref{fig2}b), a monotonic increase (Fig.~\ref{fig2}c), and those following other trends such as non-monotonic or modest band-gap variation (Fig.~\ref{fig2}d). It is important to emphasize that the maximum distortion amplitude of $0.4$~\AA~ has been arbitrarily chosen and that, while numerically plausible, may not correspond to actual thermal displacements. We will revisit and discuss this important caveat in the next section.

In Fig.~\ref{fig2}b, the chalcohalide anti-perovskites show a pronounced reduction in band gap under polar phonon distortion, in agreement with previous results \cite{benitez2025giant}. Notably, Ag$_3$SBr and Ag$_3$SI exhibit band gap decreases of approximately 30\% at a distortion amplitude of 0.4~\AA. In contrast, compounds such as RbHO and YHO$_2$ display much weaker responses, with band gap reductions of less than 4\% at the same displacement amplitude. Although these latter materials were initially identified as promising candidates for strong electron-phonon coupling, our specific analysis indicates that their actual tunability may be limited.

Figure~\ref{fig2}c displays materials that exhibit a pronounced and steady increase in $E_{g}$ as the phonon distortion amplitude grows. All these materials adopt perovskite-like structures with $Pm\overline{3}m$ symmetry. Compounds such as KHgF$_3$ and PbTiO$_3$ show significant band gap increases of approximately 20\% for a distortion amplitude of 0.4~\AA. Remarkably, BaTiO$_3$ and NaNbO$_3$-I exhibit exceptional enhancements of nearly 100\% under the same conditions. 

Figure~\ref{fig2}d presents the remaining materials, which do not exhibit monotonic or significant phonon-induced $E_{g}$ variations. InHO$_2$ displays a mild band-gap reduction, whereas the other compounds, also perovskite-like, show only modest band-gap increases (less than 10\%) under the maximum considered phonon distortion. An especially noteworthy case is NaNbO$_3$, which demonstrates that phonon-induced band-gap variation can strongly depend on the specific polymorph. In particular, the $Pm\overline{3}m$ phase (denoted as polymorph I) shows strong electron-phonon coupling and significant band-gap modulation, while the $P4/mbm$ phase (polymorph II) exhibits only weak phonon-mediated effects. 

We do not find a correlation between the (imaginary) energy of the largest polar $\Gamma$-point phonon instability ($E_{\Gamma}$ in Table~I) and the magnitude of the associated phonon-induced band-gap variation. For example, RbHO exhibits the largest $E_{\Gamma}$ among all candidates, yet shows only minimal $E_{g}$ modulation (Fig.~\ref{fig2}b). In contrast, Sr$_{3}$CdO$_{4}$ has the smallest $E_{\Gamma}$, yet displays an appreciable change in $E_{g}$ (Fig.~\ref{fig2}c). These observations show that $E_{\Gamma}$ alone is not a reliable descriptor for identifying materials with strong band-gap tunability under polar phonon distortions.

\subsection*{Representative materials: Ag$_3$SBr, BaTiO$_3$ and PbHfO$_3$}
\label{subsec:representative}
\begin{figure*}
    \centering
    \includegraphics[width=1.0\linewidth]{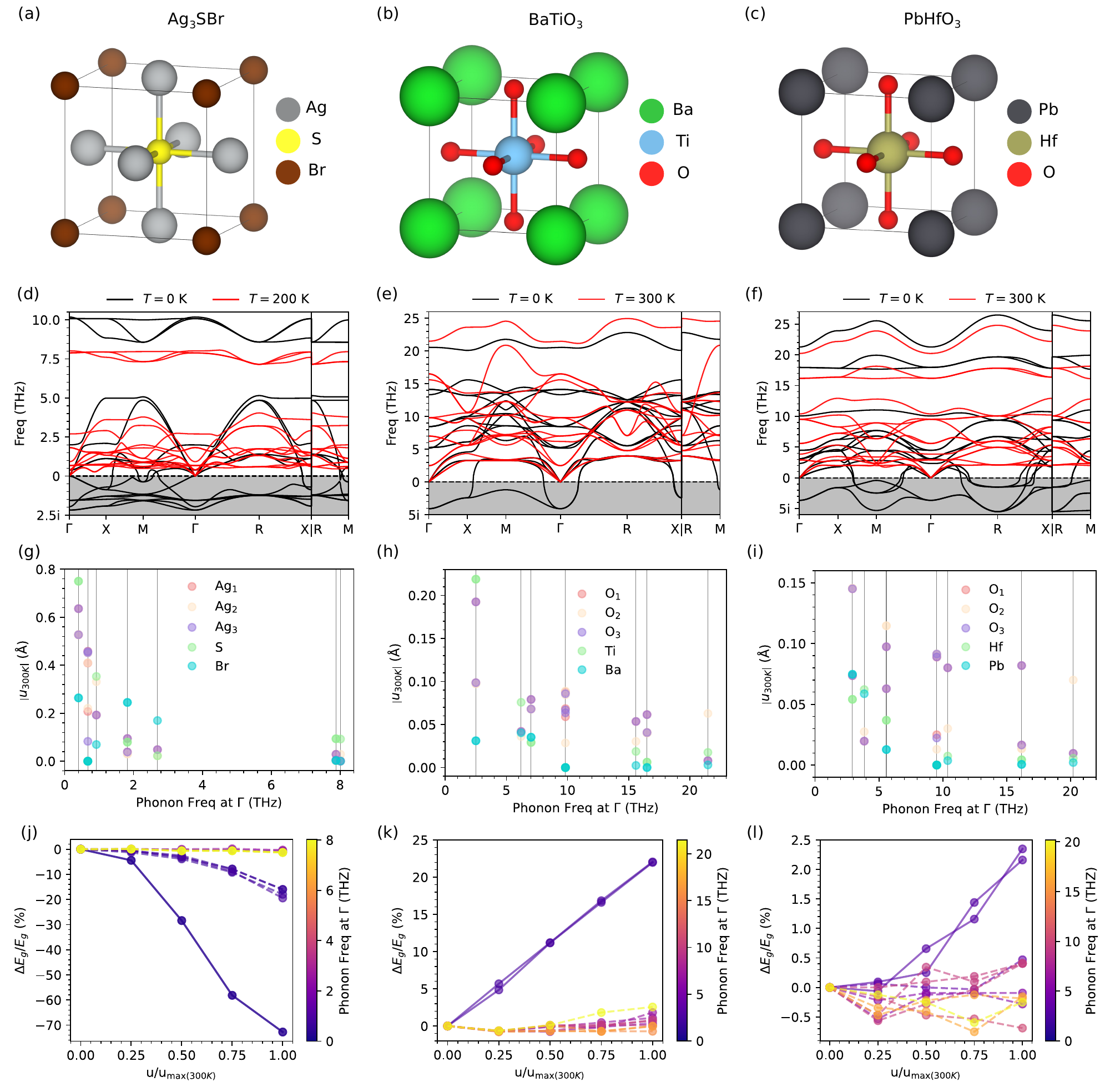}
    \caption{\textbf{Phonon dispersions and band gap relative variation for representative materials.} 
    (a--c)~Unit cell structure of Ag$_3$SBr (band-gap decreases), BaTiO$_3$ (increases), and PbHfO$_3$ (minimal change).
    (d--f)~Harmonic (black) and $T$-renormalized (red) phonon dispersions.
    (g--i)~Atomic displacements corresponding to phonon distortions calculated at finite temperature. Vertical lines indicate phonon energies at the $\Gamma$-point.
    (j--l)~Band gap changes induced by $\Gamma$-point phonon distortions, scaled by the maximum amplitude obtained at finite temperature. Lines are guides to the eye; two key low-energy optical modes are represented with solid lines.}
    \label{fig3}
\end{figure*}

We selected three representative materials for in-depth analysis, each exemplifying one of the distinct trends in band-gap behavior: Ag$_3$SBr, which exhibits a strong band gap reduction; BaTiO$_3$, which shows a pronounced band gap increase; and PbHfO$_3$, which displays minimal variation. The corresponding crystal structures are depicted in Figs.~\ref{fig3}a--c. In conventional perovskites such as BaTiO$_3$ and PbHfO$_3$, anions (i.e., oxygen atoms) occupy the corners of the octahedra, with one type of cation residing at the octahedral center and another at the cube corners. In contrast, in the anti-perovskite Ag$_3$SBr, these roles are inverted: the octahedral corners are occupied by cations (i.e., silver atoms), while anions reside at the center and corner positions.

The high-symmetry cubic $Pm\overline{3}m$ phase in perovskite-like structures is known to be vibrationally unstable at $T = 0$~K; however, it can be dynamically stabilized at finite temperatures due to thermal lattice effects \cite{rurali2024giant, cazorla2024light, wang2021finite}. In practice, this cubic polymorph is often observed at moderate and high temperatures as a result of phase transitions from lower-symmetry structures, typically involving distortions or tilting of the anion octahedra \cite{cohen1992origin, zhong1994phase}. To characterize and analyze the vibrational behavior of the $Pm\overline{3}m$ phase under realistic thermal conditions, we computed the phonon spectra of Ag$_3$SBr, BaTiO$_3$, and PbHfO$_3$ at both $T = 0$ and $300$~K (Figs.~\ref{fig3}d--f), including long-range dipole-dipole interactions (Methods). For Ag$_3$SBr, finite-temperature calculations were conducted at $200$~K to avoid inaccuracies arising from superionicity \cite{benitez2025crystal}. At zero temperature, all three compounds exhibit several optical phonon branches with imaginary frequencies, confirming their dynamic instability in this limit. However, upon incorporating thermal effects (Methods), these instabilities are removed, in agreement with experimental observations at ambient and high-$T$ conditions.

For a $\Gamma$-point phonon mode indexed by $\nu$, the maximum amplitude of the vibrational displacement for atom $j$ along the $\alpha$ Cartesian direction is given by \cite{phonopy-phono3py-JPSJ}:
\begin{equation}
    u_{j,\nu}^{\alpha}=\sqrt{\frac{\hbar}{2m_{j}\omega_{\nu}}} \sqrt{1 + 2n_{\nu}(T)}\left| \mathbf{e}_{j,\nu}^{\alpha} \right|, 
\label{eq1}
\end{equation}
where $\omega_\nu$ is the mode frequency, $m_j$ is the mass of atom $j$, $n_\nu (T)$ is the Bose-Einstein occupation factor at temperature $T$, and $\mathbf{e}_{j,\nu}^{\alpha}$ is the normalized eigenvector component of the phonon mode. From this expression, it follows that atomic displacements become larger for lighter atoms, low-frequency modes, and elevated temperatures. This behavior is clearly reflected in our results shown in Figs.~\ref{fig3}g--i, where we depict the magnitude of the atomic displacements associated with $\Gamma$-point phonons at $T = 300$~K (made the exception of Ag$_3$SBr, for which finite-temperature calculations were conducted at $200$~K to avoid inaccuracies arising from superionicity \cite{benitez2025crystal}).

Ag$_3$SBr, which features ultra low-energy optical modes around $0.4$~THz, exhibits exceptionally large phonon-induced atomic displacements: up to $0.8$~\AA~ for S atoms and $0.6$~\AA~ for Ag atoms (Fig.~\ref{fig3}g). In contrast, BaTiO$_3$ and PbHfO$_3$, whose lowest-energy phonon modes lie at approximately $2.5$ and $3.0$~THz respectively, display significantly smaller, though still considerable, displacements: $0.1$--$0.2$~\AA~ for BaTiO$_3$ (Fig.~\ref{fig3}h) and $0.05$--$0.15$~\AA~ for PbHfO$_3$ (Fig.~\ref{fig3}i). For higher-energy phonons, the atomic displacements are further suppressed, remaining below $0.2$~\AA~ in Ag$_3$SBr and below $0.1$~\AA~ and $0.05$~\AA~ in BaTiO$_3$ and PbHfO$_3$, respectively. These findings indicate that Ag$_3$SBr is likely more anharmonic and exhibits stronger electron-phonon interactions than the two perovskites, owing to its larger vibrational amplitudes and, consequently, greater potential for lattice-mediated electronic hybridization.

\begin{table*}[!htbp]
    \centering
    \begin{tabular}{ccccccccc}
    \hline
    \hline
	    & & & & & & & &   \\
	    Material \qquad & \qquad $E_{g}^{0{\rm K}}$ (eV) \qquad & \qquad $ E_{g}^{300{\rm K}}$ (eV) & $\Delta E_{g}^{S}$ (meV) & $\Delta E_{g}^{L}$ (meV) \qquad & \qquad $ E_{g}^{600{\rm K}} $ (eV) & $\Delta E_{g}^{S} $ (meV) & $\Delta E_{g}^{L} $ (meV) \qquad &  \qquad $ E_{g}^{\rm exp}$ (eV) \\
	    & & & & & & & & \\
    \hline
	    & & & & & & & & \\
	    Ag$_3$SBr \qquad & \qquad 1.8$\pm$0.1 \qquad & \qquad 1.2$\pm$0.1 & $ -480 $ & $ -74 $ \qquad & \qquad 0.9$\pm$0.2 & $ -680 $ & $ -175 $ \qquad & \qquad 1.0 \cite{cano2024novel} \\
            BaTiO$_3$  \qquad & \qquad 2.5$\pm$0.1 \qquad & \qquad 3.5$\pm$0.1 & $ +1060 $ & $ -38 $ \qquad & \qquad 3.1$\pm$0.1 & $ +770 $ & $ -160 $ \qquad & \qquad 3.2 \cite{wemple1970polarization} \\
            PbHfO$_3$ \qquad & \qquad 3.2$\pm$0.1 \qquad & \qquad 3.4$\pm$0.1 & $ +220 $ & $ -65 $ \qquad & \qquad 2.9$\pm$0.1 & $ -50 $ & $ -230 $ \qquad & \qquad 3.4 \cite{zhang2012electronic}  \\
	    & & & & & & & &  \\
    \hline
    \hline
    \end{tabular}
	\caption{{\bf Temperature-dependence of the calculated band gaps.} 
	$E_{g}$ values were obtained at zero temperature (excluding quantum nuclear effects), $300$~K, and $600$~K. Short- and long-wavelength phonon band-gap corrections, $\Delta E_{g}^{S}$ and $\Delta E_{g}^{L}$ (Methods), are provided at each temperature. Numerical uncertainties are provided, mainly resulting from the $\Delta E_{g}^{S}$ correction term. Experimental room-temperature band gaps for Ag$_3$SBr \cite{cano2024novel}, BaTiO$_3$ \cite{wemple1970polarization} and PbHfO$_3$ \cite{zhang2012electronic}, are shown for comparison.}
    \label{tab:bandgap}
\end{table*}

\begin{figure*}
    \centering
    \includegraphics[width=1.0\linewidth]{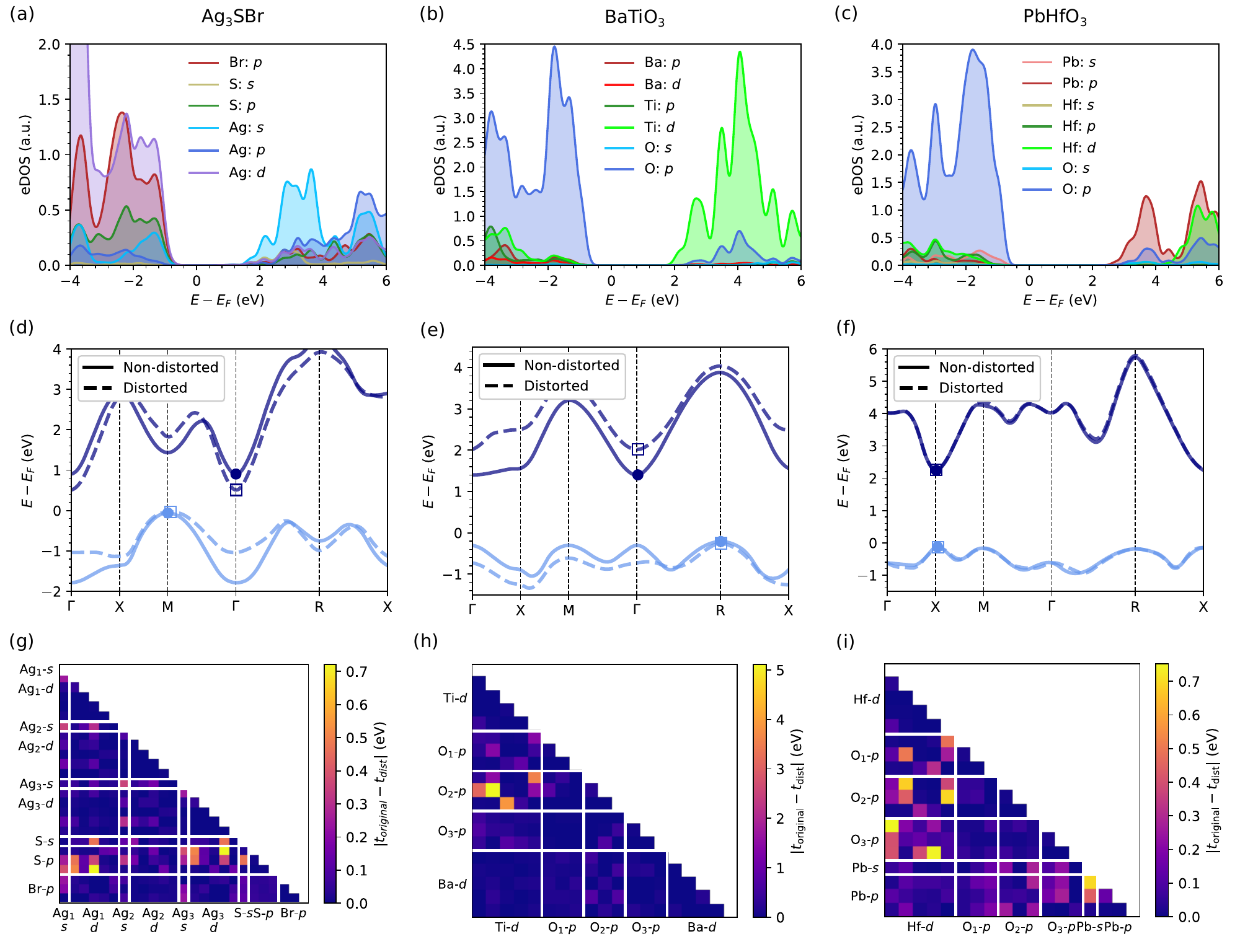}
    \caption{\textbf{Electronic properties and band-gap variation mechanisms for representative materials.} 
	(a--c)~Density of states near the band gap for Ag$_3$SBr, BaTiO$_3$, and PbHfO$_3$, showing contributions from the most relevant orbitals.
	(d--f)~Valence band (light blue) and conduction band (dark blue) for the equilibrium (solid) and phonon distorted (dashed) structures.
	(g--i)~Changes in tight-binding hopping terms for each compound under phonon distortion. Warmer colors indicate greater changes in orbital overlap; white lines separate orbital types. The Hamiltonian matrix is symmetric.}
    \label{fig4}
\end{figure*}

Figures~\ref{fig3}j--l present the relative band gap variation with respect to the equilibrium (undistorted) structure, using phonon-mode displacements with maximum amplitudes computed at $T = 300$~K via Eq.~(\ref{eq1}). In Ag$_3$SBr (Fig.~\ref{fig3}j) and BaTiO$_3$ (Fig.~\ref{fig3}k), we observe that low-energy phonon modes are primarily responsible for substantial band gap changes: up to a $70$\% reduction in Ag$_3$SBr and more than a $20$\% increase in BaTiO$_3$ at the corresponding maximum displacements. In Ag$_3$SBr, low-frequency lattice vibrations strongly reduce the band gap, whereas higher-frequency modes have a negligible effect. In BaTiO$_3$, low-energy modes markedly increase the band gap, while higher-energy modes produce minor changes, with some even slightly decreasing the gap. The smaller size of the band gap enhancement in BaTiO$_3$ compared to that shown in Fig.~\ref{fig2}c arises from the reduced phonon displacement amplitudes calculated at $T = 300$~K, which are all smaller than $0.4$~\AA.

It is important to note that as temperature increases, the relative influence of phonon modes changes due to the Bose-Einstein occupancy distribution: low-energy phonons become less dominant, while higher-energy modes gain increasing weight (Supplementary Fig.~1 and Supplementary Discussion). Consequently, even if low-energy phonons initially drive a band gap increase or decrease at low temperatures, a compensating, or even opposite, effect may emerge at higher temperatures, leading to $T$-induced non-monotonic band gap behavior. Examples of semiconductors where the band gap first increases with temperature before decreasing include chalcopyrites \cite{bhosale2012temperature} and single-walled carbon nanotubes \cite{lefebvre2004temperature}.

In the case of PbHfO$_3$ (Fig.~\ref{fig3}l), phonon-mode distortions generally lead to minimal changes in the band gap, many of which fall within the typical accuracy limit of our DFT calculations ($\sim 0.1$~eV). Nevertheless, the low-energy modes exhibit a consistent, albeit modest, tendency to increase the band gap by approximately $2.5$\% at $300$~K. In contrast, certain higher-energy modes exert a weak decreasing effect. This indicates a temperature-driven competition between phonon modes that slightly increase and others that slightly decrease the band gap. As a result, even though PbHfO$_3$ may undergo sizable atomic displacements due to anharmonicity and thermal effects, its net band-gap variation remains small. As we will discuss later, this limited tunability is closely linked to the specific nature of electronic orbital hybridizations in this compound.

\subsection*{Band-gap dependence on temperature}
\label{subsec:T-dependence}
We computed the thermally renormalized band gaps of the three representative materials, Ag$_3$SBr, BaTiO$_3$ and PbHfO$_3$, using the methodology described in the Methods section and in work \cite{benitez2025giant}. Calculations were performed at $T = 300$ and $600$~K, accounting for both short-range and long-range electron-phonon contributions. Long-range contributions, which stem from limitations associated with the use of finite supercell sizes in simulations of polar materials, were corrected using the Fr\"ohlich polaron approach \cite{zacharias2020fully, chen2024temperature} (Methods). Short-range contributions, were corrected using finite-difference methods designed to handle the strong anharmonicity (i.e., imaginary phonon frequencies) present in these systems \cite{monserrat2018electron} (Methods). Quantum nuclear effects \cite{cazorla2017simulation} were systematically neglected, as they are expected to be small in materials composed of heavy atoms and their exclusion significantly improves computational efficiency. Our results are summarized in Table~II and compared with available experimental room-temperature data from the literature.

We found excellent agreement between our calculated finite-temperature band gaps and experimental room-temperature data for all three materials considered (Table~II). For both Ag$_3$SBr and BaTiO$_3$, there exists a substantial discrepancy of approximately $0.8$~eV between the computed zero-temperature band gaps and the corresponding experimental room-temperature values: the theoretical $E_g$ of Ag$_3$SBr is considerably larger than the experimental one, whereas that of BaTiO$_3$ is significantly smaller. However, when the band gaps are computed at $T = 300$ and $600$~K, we observe a pronounced reduction in $E_g$ for Ag$_3$SBr and an increase for BaTiO$_3$, leading to excellent agreement with the experimental data within the numerical uncertainties. In the case of PbHfO$_3$, the zero-temperature band gap is already in good agreement with the experiments, with a deviation of only $\sim 0.1$~eV. 

For Ag$_3$SBr, both short- and long-range electron-phonon interactions contribute to the reduction of the band gap. While short-range effects dominate, being approximately $4$--$6$ times larger than the long-range contributions, the latter become increasingly significant as temperature rises. At $T = 300$~K, the band gap is reduced by approximately $33$\% relative to its zero-temperature value, indicating a giant thermal renormalization effect, consistent with the findings reported in Ref.~\cite{benitez2025giant}. At $T = 600$~K, the band gap is further reduced to nearly half of its static value, underscoring the pronounced role of electron-phonon coupling in determining the finite-temperature electronic properties of this material.

For BaTiO$_3$, we observe a pronounced band gap increase primarily driven by short-range phonon contributions. At $T = 300$~K, the short-range correction enhances the band gap by nearly $1.0$~eV. At $T = 600$~K, this positive contribution slightly diminishes, yielding an increase of approximately $0.8$~eV relative to the zero-temperature value. As mentioned earlier, as temperature rises higher-energy phonon modes gain importance and, since these modes tend to reduce the band gap (Fig.~\ref{fig3}k), the overall temperature dependence is non-monotonic. The long-range Fröhlich contribution is comparatively minor, resulting in a consistent band gap reduction of $0.04$~eV at $T = 300$~K and $0.2$~eV at $600$~K.

For PbHfO$_3$, the long-range phonon contribution surpasses the short-range component at elevated temperatures, resulting in a modest overall reduction of the band gap (Fig.~\ref{fig3}l). At $T = 300$~K, short-range interactions increase the band gap by approximately $0.2$~eV, whereas long-range effects reduce it by $0.07$~eV. At $T = 600$~K, the total variation in the band gap remains below $10$\%, driven by a more pronounced long-range reduction of $0.2$~eV and a concurrent, though smaller, short-range decrease of $0.05$~eV. 

From this analysis, we conclude that Ag$_3$SBr and BaTiO$_3$ exhibit pronounced electron-phonon interactions that significantly influence their optoelectronic properties. As such, incorporating these effects is crucial for achieving accurate agreement with experimental measurements. In contrast, despite the strong anharmonicity of PbHfO$_3$, electron-phonon interactions exert only a limited effect on its band gap. It is important to underscore that most band gap calculations reported in the literature are performed at $T = 0$~K, often neglecting thermal effects under the assumption that they are negligible. However, our results demonstrate that this assumption does not hold universally. Accurately capturing the temperature dependence of optoelectronic properties in semiconductors, especially in materials with soft phonon modes or strong lattice anharmonicity, requires explicit consideration of electron-phonon interactions.

\subsection*{Electron-phonon coupling mechanisms}
\label{subsec:e-phon}
To elucidate why Ag$_3$SBr and BaTiO$_3$ exhibit strong but opposite band gap trends with increasing temperature, while PbHfO$_3$ does not, we further analyzed their underlying electronic mechanisms using complementary tight-binding (TB) models. Our analysis focuses on the low-energy polar optical phonon modes, which dominate electron-phonon interactions, particularly at low temperatures.

Figures~\ref{fig4}a--c display the electronic density of states (eDOS) near the valence band maximum (VBM) and conduction band minimum (CBM) for the three representative materials, as computed with DFT. In Ag$_3$SBr, the valence band comprises a mixture of Ag-$d$, Br-$p$, S-$p$, and Ag-$s$ orbitals. In contrast, the valence bands of the perovskites BaTiO$_3$ and PbHfO$_3$ are primarily dominated by O-$p$ states. Regarding the conduction band, Ag$_3$SBr features contributions from Ag-$s$, S-$s$, and Ag-$d$ orbitals; BaTiO$_3$ mainly from Ti-$d$ and O-$p$; and PbHfO$_3$ from Pb-$p$ and O-$p$. These band-edge orbitals are crucial for understanding the distinct thermal band-gap behaviours observed in these materials, as we elaborate next.

Figures~\ref{fig4}d--f display the band structures calculated near the VBM and CBM for the three materials in both their equilibrium (undistorted, solid lines) and phonon-distorted (dashed lines) configurations. The distorted structures correspond to atomic displacements of $0.2$~\AA~ along the eigenvector of the lowest-energy optical phonon mode at the $\Gamma$-point. To reduce computational cost, these band structures were computed using a standard semi-local exchange-correlation functional. This approximation is justified because, while the absolute band gap values may be underestimated, the fundamental band-dispersion trends are well reproduced (Supplementary Fig.~2). 

In Ag$_3$SBr, we observe that the CBM shifts to lower energies under the phonon distortion, while the VBM remains essentially unchanged and becomes slightly flatter, in agreement with prior work \cite{benitez2025giant}. The observed band gap reduction in Ag$_3$SBr is therefore primarily driven by the downward shift of the CBM. In BaTiO$_3$, the CBM shifts upward noticeably, while the VBM slightly moves downward. Consequently, the band gap increase in BaTiO$_3$ results from both an upward shift of the CBM and a downward shift of the VBM. In contrast, PbHfO$_3$ exhibits minimal changes, with only a barely detectable downward shift of the VBM, consistent with the small band gap increase reported in Fig.~\ref{fig3}l.

To gain deeper insight into the electronic mechanisms driving these trends, we employed a TB model (Methods) to analyze how orbital hybridizations evolve under phonon distortion. Figures~\ref{fig4}g--i show the changes in TB hopping parameters between different atomic orbitals before and after introducing the phonon distortion. In the TB formalism, the hopping parameters, which quantify the probability amplitude for an electron to \textit{hop} from one atomic orbital to another, are given by the off-diagonal elements of the TB Hamiltonian, $\left< n \right| H \left| m \right>$, where $n$ and $m$ denote the involved electronic orbitals, and $H$ is the TB Hamiltonian. The diagonal terms of the Hamiltonian, $\left< n \right| H \left| n \right>$, provide the kinetic energy of the orbitals. When two orbitals hybridize, both a larger hopping term and a smaller kinetic energy difference between them are associated with an increased energy splitting for the resulting bonding and antibonding states.

In Ag$_3$SBr, the most notable, albeit still modest, changes in electronic hybridization occur between the S-$p$ and Ag-$d$ orbitals (Fig.~\ref{fig4}g). However, this type of orbitals hybridization does not contribute significantly to the CBM and therefore cannot account for the observed band gap trends. In contrast, we observe substantial changes in the difference between the kinetic energies of the Ag-$s$ and S-$s$ orbitals before and after introducing the distortion. Supplementary Table II presents the numerical values for the relevant orbitals, showing that the hopping term increases slightly by $0.13$ eV with the distortion, while the difference in kinetic energies between the Ag-$s$ and S-$s$ orbitals decreases by a total of $1.63$ eV. The notable decrease in the difference of the diagonal terms leads to a pronounced lowering of the energy of the bonding states formed by Ag-$s$ and S-$s$ hybridization. Since these bonding states contribute significantly to the CBM, their energy reduction results in a substantial narrowing of the band gap \cite{benitez2025giant}.

In BaTiO$_3$, the most significant changes in hybridization are observed between the O-$p$ and Ti-$d$ orbitals (Fig.~\ref{fig4}h). Phonon distortions notably increase the energy splitting between the bonding and antibonding states arising from this orbital hybridization. From Supplementary Table II, we observe a substantial increase of $5.11$ eV in the hopping term between the O-$p$ and Ti-$d$ orbitals. There is also a decrease in the difference between their kinetic energies, although this is modest as it amounts to $0.52$ eV. Since the bonding states lie at the VBM and the antibonding states at the CBM, this enhanced splitting leads to an opening of the band gap.

In PbHfO$_3$, the primary hybridization change involves the O-$p$ and Hf-$d$ orbitals (Fig.~\ref{fig4}i). Although a bonding-antibonding splitting occurs, similarly to BaTiO$_3$, the antibonding states lie at energies well above the CBM, and therefore do not significantly contribute to band gap variation. The slight upward shift in the VBM, driven by the O-$p$ and Hf-$d$ orbital hybridization, accounts for the modest band gap increase. Furthermore, the changes in hopping parameters are smaller than in BaTiO$_3$, resulting in weaker orbital splitting and consequently a smaller 
overall effect on the band gap.

\begin{figure*}
    \centering
    \includegraphics[width=0.9\linewidth]{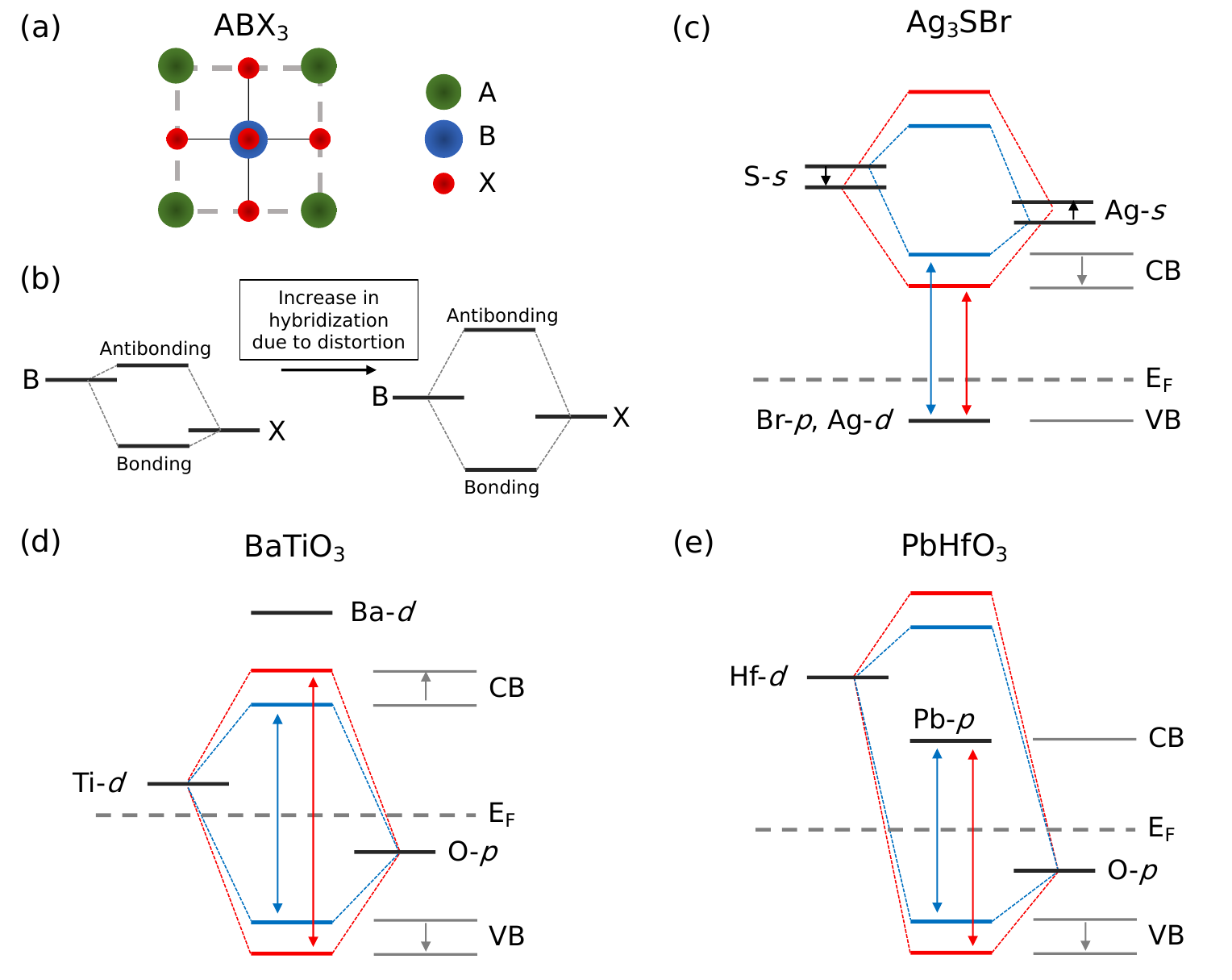}
    \caption{\textbf{General electron-phonon coupling mechanisms driving temperature-induced band gap variation in 
	perovskite-like semiconductors.} 
	(a)~Equilibrium perovskite-like ABX$_3$ structure.
	(b)~Increased bonding-antibonding splitting from orbital hybridization due to a soft polar phonon mode distorsion.
	(c)~Band-gap reduction in Ag$_3$SBr resulting from Ag--S $s$-orbitals hybridization and conduction band lowering.
	Band gap increase via hybridization in (d)~BaTiO$_3$ and (e)~PbHfO$_3$. In BaTiO$_3$, bonding and antibonding states 
	lie at the VBM and CBM, respectively. In PbHfO$_3$, the antibonding state lies above the CBM hence phonon-induced 
	splitting causes an almost negligible band-gap change.}
    \label{fig5}
\end{figure*}

The concept that band gap variations can arise from the alignment of the valence and conduction bands with bonding and antibonding states and from changes in their energy splitting, was previously investigated in Ref.~\cite{ahmed2022luminescence}. However, while that study focused on how chemical doping in perovskite-like systems modifies orbital hybridization, and consequently the band gap, our work demonstrates that similar hybridization changes can also be induced by phonon distortions, specifically those associated with low-energy polar modes.

\subsection*{General electron-phonon coupling framework}
\label{subsec:general}
To generalize and summarize the electronic behavior observed in Ag$_3$SBr, BaTiO$_3$, and PbHfO$_3$, we focus on the archetypal non-distorted perovskite-like structure with symmetry group $Pm\overline{3}m$, depicted in Fig.~\ref{fig5}a. In this structure, the atoms labeled B (center of the octahedron) and X (octahedral corners) are crucial in the orbital hybridizations that influence band gap variations.

Figure~\ref{fig5}b illustrates the general principle: hybridization between the B and X atom orbitals results in bonding and antibonding states. When the lattice is distorted by a polar phonon mode, the increased orbital overlap and the reduction between their difference in kinetic energies enhance the splitting between these states. Whether this splitting leads to an increase or decrease in the band gap depends on the relative positions of the Fermi level and bonding/antibonding states, as well as the presence of other orbital contributions within the hybridization splitting. This framework enables us to rationalize all potential cases of band gap variation under low-energy phonon distortions in centrosymmetric perovskite-like materials. 

Figures~\ref{fig5}c--e schematically represent the orbitals hybridization scenarios corresponding to the three materials analyzed in previous sections. The changes in kinetic energy are only illustrated in Fig. 5c because they are primarily responsible for the decrease in the bonding state energy of this material. We do not show them for the other two materials, as their contributions are less significant. Numerical values of the relevant TB matrix elements are provided in Supplementary Table II.

Based solely on the eDOS of a given material, now we can explain why the different perovskite-like systems reported in Fig.~\ref{fig2} exhibit distinct phonon-induced band gap behaviors. Supplementary Fig.~3 presents the eDOS of the equilibrium structures for additional perovskite-like compounds identified in our screening, all with $Pm\overline{3}m$ symmetry. These include Ag$_3$SeBr, Ag$_3$SeI, Ag$_3$SI, PbTiO$_3$, NaNbO$_3$, KNbO$_3$, NaPaO$_3$, AgPaO$_3$, KHgF$_3$, and RbPbF$_3$. For the anti-perovskite compounds reported in Fig.~\ref{fig2}, we anticipate a similar band gap reduction mechanism as observed in Ag$_3$SBr, given their isoelectronic nature and eDOS similarity (Supplementary Figs.~3a--c).

For PbTiO$_3$, NaNbO$_3$, and KNbO$_3$ (Supplementary Figs.~3d--f), the scenario closely resembles that of BaTiO$_3$. Their VBM are dominated by O-$p$ orbitals, while the CBM primarily consist of B-site $d$ electrons. Consequently, phonon-induced distortions are expected to enhance the hybridization between the B-$d$ and O-$p$ orbitals, resulting in a widening of the band gap. This aligns with the positive band gap shifts reported in Fig.~\ref{fig2}. Likewise, NaPaO$_3$ and AgPaO$_3$ exhibit an eDOS similar to that of BaTiO$_3$, although in these systems the relevant orbital hybridizations occur between O-$p$ and Pa-$f$ states.

For KHgF$_3$ and RbPbF$_3$ (Supplementary Figs.~3i--j), the band gap increase can also be interpreted using the hybridization mechanism discussed for BaTiO$_3$. In both materials, the valence and conduction bands are primarily composed of X-site (F) and B-site (Hg or Pb) orbitals, likely forming bonding and antibonding states. Upon phonon distortion, the increased splitting of these states results in a larger band gap.

\section*{Discussion}
\label{sec:discussion}
We have demonstrated that electron-phonon corrections to the band gap in perovskite-like systems can be effectively explained through a combination of phononic and electronic coupled mechanisms. The band gap change is directly linked to distortions induced by polar lattice vibrations, which alter the hybridization between electronic orbitals. This hybridization reshapes the energy landscape of the electronic bands, and when these changes impact the extrema of the valence and conduction bands, the band gap is modified. A central factor in this optoelectronic modulation mechanism is anharmonicity, which is associated with low-energy optical modes capable of producing large atomic displacements. Nevertheless, as illustrated by PbHfO$_3$, anharmonicity alone does not guarantee strong band gap renormalization; effective coupling between specific phonon modes and electronic states near the band edges is essential.

The polar character of the phonon modes under study is particularly significant, as it implies they can be externally excited using electric fields, rather than relying solely on thermal activation. This opens the door to potential applications in optoelectronic devices where the band gap, and more broadly the electronic and optical properties, can be dynamically tuned with electric fields. In this work, we focused primarily on band gap changes, however electron-phonon interactions can influence the entire electronic structure and, consequently, the optical response of the material. For instance, previous research on chalcohalide anti-perovskites has demonstrated a significant increase in the absorption coefficient with temperature, making these materials promising for photovoltaic applications \cite{benitez2025giant}. Low-energy optical non-polar modes may similarly lead to significant electron-phonon effects, although since they cannot be excited via electric fields were neglected in this study. Nevertheless, with advances in laser technology and time-resolved spectroscopy, it is now feasible to excite non-polar modes using light sources \cite{kennes2017transient,dekorsky1993coherent,pomarico2017enhanced}, broadening the range of tools available for efficient control or functionality.

From our screening, perovskite-like systems emerge as the most promising materials for exhibiting large optoelectronic tunability. However, other material families not covered by our high-throughput study may also be of interest. Exploring these possibilities will require broader computational screenings that extend beyond the $\Gamma$-point, including phonon contributions throughout the entire Brillouin zone, and additional materials to those contained in the PhononDB database \cite{phonondb}. Fortunately, the rapid development of machine learning interatomic potentials (MLIPs) \cite{jacobs2025practical}, such as M3GNet \cite{chen2022universal} and MACE \cite{batatia2023foundation}, offers a promising path forward in high-throughput modeling of material vibrational properties. These methods enable accurate force calculations and molecular dynamics simulations at a fraction of the cost of DFT, with increasing precision \cite{lee2025accelerating}. 

Finally, solid solutions such as Ag$_3$SBr$_x$I$_{1-x}$ or Na$_x$K$_{1-x}$NbO$_3$ are also expected to exhibit significant electron-phonon effects and phononic behavior due to their similarities in structural and optoelectronic properties to their parent compounds \cite{cano2024novel}. Considering alloyed compounds dramatically expands the landscape of potentially interesting materials for tunable optoelectronic applications. Addressing the complexity of such systems will greatly benefit from MLIP-based techniques and machine learning tools such as graph neural networks \cite{xie2018crystal}. The authors are actively pursuing this computational materials research direction.

\section*{Conclusions}
\label{sec:conclusions}
From our screening of approximately $10,000$ materials, we identified several hundred candidates with significant band gap changes driven by low-energy polar phonon modes, enabled by strong electron-phonon coupling. We validated this behavior in a subset of perovskite-like systems, presenting and generalizing both phononic and electronic mechanisms to explain the diverse electron-phonon renormalization effects found on the materials optoelectronic properties. Additionally, we supported our theoretical findings for temperature-induced band gap variations with experimental evidence, confirming the accuracy of our approach and explaining possible discrepancies between ambient experimental results and zero-temperature computational predictions.

This work not only advances the fundamental understanding of electron-phonon interactions in perovskite-like materials, through clear and intuitive physical and chemical reasoning, it also lays the theoretical groundwork for leveraging these strong renormalization effects in future technological applications. These findings open the possibility of dynamically tuning the optoelectronic properties of semiconductors using electric fields, temperature, or light, offering exciting opportunities for next-generation optoelectronic devices.

\section*{Methods}
\label{sec:methods}
{\bf Zero-temperature first-principles calculations.}~DFT calculations \cite{blochl1994projector,cazorla2017simulation} were performed with the \verb!VASP! software \cite{kresse1993ab,kresse1996efficiency,kresse1996efficient} and semilocal PBEsol exchange-correlation functional \cite{perdew2008restoring}. Wave functions were represented in a plane-wave bases set truncated at $700$~eV. We selected a dense $\mathbf{k}$-point grid, with $8 \times 8 \times 8$ points for the reciprocal-space Brillouin zone (BZ) sampling, for the cubic perovskite systems. We obtained zero-temperature energies converged to within $0.5$~meV per formula unit. For geometry relaxations, a force tolerance of $0.005$~eV\,\AA$^{-1}$ was imposed in all the atoms. The electronic bands and electronic density of states were estimated using the hybrid functional HSEsol and considering spin-orbit coupling effects \cite{schimka2011improved,krukau2006influence}. This level of theory is usually enough to get reliable electronic results \cite{garza2016predicting,ganose2016interplay}. Quantum nuclear effects were disregarded throughout this work.
\\

{\bf Finite-temperature first-principles simulations.}~\emph{Ab initio} molecular dynamics (AIMD) simulations were performed in the canonical $NVT$ ensemble, neglecting thermal expansion effects and employing two different simulation cells containing $40$ and $320$ atoms with periodic boundary conditions applied along the three Cartesian directions. The temperature in the AIMD simulations was kept fluctuating around a set-point value by using Nose-Hoover thermostats \cite{nose1984unified,hoover1985canonical}. Newton's equations of motion were integrated using the standard Verlet's algorithm with a time step of $1.5 \cdot 10^{-3}$~ps. $\Gamma$-point sampling for reciprocal-space integration was employed in the AIMD simulations, which spanned approximately over $100$~ps. These calculations were performed with the semilocal PBEsol exchange-correlation functional \cite{perdew2008restoring}.
\\

{\bf Harmonic phonon calculations.}~The second-order interatomic force constant matrix for the three selected materials and resulting harmonic phonon spectrum were calculated with the finite-differences method as is implemented in the \verb!PhonoPy! software \cite{phonopy-phono3py-JPCM,phonopy-phono3py-JPSJ}. $2 \times 2 \times 2$ and $4 \times 4 \times 4$ supercells with a dense $\mathbf{k}$-point grid of $4 \times 4 \times 4$ and $2 \times 2 \times 2$ for BZ sampling, respectively, were employed for the phonon calculations of targeted structures. These calculations were performed with the semilocal PBEsol exchange-correlation functional \cite{perdew2008restoring}. The non-analytical term for polar materials \cite{pick1970microscopic} was taken into consideration through Gonze's method \cite{gonze1994interatomic} using the Born effective charges and dielectric tensor. 
\\

{\bf Anharmonic phonon calculations.}~The \verb!DynaPhopy! software \cite{carreras2017dynaphopy} was used to calculate the anharmonic lattice dynamics (i.e., $T$-renormalized phonons) of the three selected materials from AIMD simulations. The supercells and simulation technical parameters described above were used in these calculations. 

A normal-mode-decomposition technique \cite{sun2014dynamic} was employed in which the atomic velocities $\mathbf{v}_{jl}(t)$ ($j$ and $l$ represent particle and Cartesian direction indexes) generated during fixed-temperature AIMD simulation runs were expressed like:
\begin{equation}
\textbf{v}_{jl} (t) = \frac{1}{\sqrt{N m_{j}}} \sum_{\textbf{q}s}\textbf{e}_{j}(\textbf{q},s) 
	e^{i \textbf{q} \textbf{R}_{jl}^{0}} v_{\textbf{q}s}(t)~,
\label{eq2}
\end{equation}
where $N$ is the number of particles, $m_{j}$ the mass of particle $j$, $\mathbf{e}_{j}(\mathbf{q},s)$ a phonon mode eigenvector ($\mathbf{q}$ and $s$ stand for the wave vector and phonon branch), $\mathbf{R}_{jl}^{0}$ the equilibrium position of particle $j$, and $v_{\mathbf{q}s}$ the velocity of the corresponding phonon quasiparticle. 

The Fourier transform of the autocorrelation function of $v_{\mathbf{q}s}$ was then calculated, yielding the power spectrum:
\begin{equation}
G_{\textbf{q}s} (\omega) = 2 \int_{-\infty}^{\infty} \langle v_{\textbf{q}s}^{*}(0) v_{\textbf{q}s}(t) \rangle e^{i \omega t} dt~. 
\label{eq3}
\end{equation}
Finally, this power spectrum was approximated by a Lorentzian function of the form:
\begin{equation}
G_{\textbf{q}s} (\omega) \approx \frac{\langle |v_{\textbf{q}s}|^{2} \rangle}{\frac{1}{2} \gamma_{\textbf{q}s} 
        \pi \left[ 1 + \left( \frac{\omega - \omega_{\textbf{q}s}}{\frac{1}{2}\gamma_{\textbf{q}s}} \right)^{2}  \right]}~, 
\label{eq4}
\end{equation}
from which a $T$-renormalized quasiparticle phonon frequency, $\omega_{\textbf{q}s} (T)$, was determined as the peak position, and the corresponding phonon linewidth, $\gamma_{\mathbf{q}s} (T)$, as the full width at half maximum. These calculations were performed with the semilocal PBEsol exchange-correlation functional \cite{perdew2008restoring}. The non-analytical term for polar materials \cite{pick1970microscopic} was taken into consideration through Gonze's method \cite{gonze1994interatomic} using the Born effective charges and dielectric tensor.
\\

{\bf Short-wavelength phonon band-gap correction.}~The electron-phonon correction to the band gap due to the short-range phonon modes was computed as the difference between the band gap at zero temperature for the static structure and the average band gap obtained from AIMD simulations performed with a supercell, namely:
\begin{equation}
    \Delta E_{g}^{S} (T) = \lim_{t_{0} \to \infty}\frac{1}{t_{0}}\int_{0}^{t_{0}}E_{g}^{\mathbf{R}(t)}dt - E_{g}(0),
\label{eq5}
\end{equation}
where $\mathbf{R}$ represents the positions of the atoms in the supercell at a given time $t$ of the AIMD simulation. This 
expression can be numerically approximated as:
\begin{equation}
 	\Delta E_{g}^{S} (T) = \frac{1}{N} \sum_{k=1}^{N} E_{g} (\lbrace {\bf R}_{k} (T) \rbrace) - E_{g} (0),
\label{eq6}
\end{equation}
where the band gap is averaged over a finite number, $N$, of configurations, as described in \cite{zacharias2020fully}. Similarly, thermal effects on the dielectric tensor were computed.

These calculations were performed with the hybrid HSEsol exchange-correlation functional and considering spin-orbit coupling effects \cite{schimka2011improved,krukau2006influence}. Due to involved high computational expense, the total number of configurations used for the average was $N = 10$ for each material and temperature. These values were found to be appropriate for obtaining band-gap results accurate to within $0.1$~eV, as described in work \cite{benitez2025giant}.
\\

{\bf Long-wavelength phonon band-gap correction.}~The electron-phonon correction to the band gap due to long-range phonon modes was computed using the Fr\"ohlich equation for a three-dimensional polar material \cite{zacharias2020fully,chen2024temperature,ponce2015temperature,zacharias2020theory}. This correction was determined as the difference in the shifts of the conduction and valence bands:
\begin{equation}
	\Delta E_{g}^{L} (T) = \Delta \epsilon_{\rm CB}^{L} (T) - \Delta \epsilon_{\rm VB}^{L} (T),  
\label{eq7}
\end{equation}
where $\epsilon_{\rm VB}$ and $\epsilon_{\rm CB}$ denote the valence and conduction band, respectively. 

The shift of each band was computed using the expression:
\begin{equation}
	\Delta \epsilon_{\rm i}^{L} (T) = \frac{2 \alpha_{P}}{\pi} \hbar \omega_{\rm LO} \tan^{-1} \left( 
	\frac{q_{\rm F}}{q_{\rm LO,i}} \right) \left[ 2n_{T} + 1 \right],
\label{eq8}
\end{equation}
where $\alpha_{P}$ represents the polaron constant, $\omega_{\rm LO}$ the phonon frequency averaged over the three longitudinal optical $\Gamma$ phonon modes \cite{de2023high}, and $q_{\rm F}$ a truncation factor that can be approximated as Debye sphere radius. $q_{\rm LO,i}$ is defined as $\sqrt{2m^{*} \left( \omega_{\rm LO} + \omega_{i} \right) / \hbar}$, $m^{*}$ being the charge carrier effective mass and $\hbar \omega_i$ the state energy. The term $n_{T}$ is the Bose-Einstein occupation number corresponding to the average LO vibrational frequency, and the polaron constant can be computed as:
\begin{equation}
\alpha_{P} = \frac{e^{2}}{4\pi \epsilon_{0}\hbar}\left(\frac{1}{\varepsilon_{\infty}}-\frac{1}{\varepsilon_{0}}\right)\left( \frac{m^{*}}{2\hbar \omega_{\text{LO}}} \right)^{1/2},
\label{eq9}
\end{equation}
where $\varepsilon_{\infty}$ is the high-frequency dielectric constant and $\varepsilon_{0}$ the static permittivity of the system. The physical quantities entering this latter expression were determined with DFT methods (for additional details see \cite{benitez2025giant}).
\\

{\bf \textit{Ab initio} tight-binding models.}~All-electron DFT calculations for \textit{ab initio} tight binding models were performed with the \verb!WIEN2K! software \cite{blaha2020wien2k} using the local-density approximation \cite{perdew1992accurate} to the exchange correlation energy along with the linearized augmented plane wave method (FP-LAPW) \cite{singh2006planewaves,blaha1990full}. The technical parameters for these calculations were a $10 \times 10 \times 10$ ${\bf k}$-point grid and a muffin-tin radius equal to $R_{\rm MT} = 7.0 / K_{\rm max}$, where $K_{\rm max}$ represents the plane-wave cutoff. Localized energy-resolved Wannier states \cite{marzari1997maximally} were then obtained for the tight-binding calculations \cite{ku2002insulating,yin2006orbital,jiang2023variation} considering the relevant Hilbert space in the interval $-10 \le E \le 20$~eV around the Fermi energy.
\\

\section*{Acknowledgments}
P.B. acknowledges support from the predoctoral program AGAUR-FI ajuts (2024 FI-1 00070) Joan Oró, which is backed by the Secretariat of Universities and Research of the Department of Research and Universities of the Generalitat of Catalonia, as well as the European Social Plus Fund. C.L. acknowledges support from the Spanish Ministry of Science, Innovation and Universities under a FPU grant. C.C. acknowledges support by MICIN/AEI/10.13039/501100011033 and ERDF/EU under the grants TED2021-130265B-C22, TED2021-130265B-C21, PID2023-146623NB-I00, PID2023-147469NB-C21 and RYC2018-024947-I and by the Generalitat de Catalunya under the grants 2021SGR-00343, 2021SGR-01519 and 2021SGR-01411. 
R.J. and B.M. acknowledge financial support from a UKRI Future Leaders Fellowship [MR/V023926/1] and S. C. and B. M. are supported by a EPSRC grant [EP/V062654/1]. B.M. also acknowledges financial support from the Gianna Angelopoulos Programme for Science, Technology, and Innovation.
Computational support was provided by the Red Española de Supercomputación under the grants FI-2024-1-0005, FI-2024-2-0003, FI-2024-3-0004,FI-2024-1-0025, FI-2024-2-0006, and FI-2025-1-0015. This work is part of the Maria de Maeztu Units of ExcellenceProgramme CEX2023-001300-M funded by MCIN/AEI (10.13039/501100011033). E.S. acknowledges the European Union H2020 Framework Program SENSATE project: Low dimensional semiconductors for optically tuneable solar harvesters (grantagreement Number 866018), Renew-PV European COST action (CA21148) and the Spanish Ministry of Science and InnovationACT-FAST (PCI2023-145971-2). E.S. and J.-Ll.T. are grateful to the ICREA Academia program.

\bibliographystyle{unsrt}

\end{document}